\documentclass[a4paper,11pt]{article}
\pdfoutput=1 
\usepackage{jinstpub} 
\usepackage{natbib} 
\usepackage{lineno} 
\title{\boldmath A study of muon-electron elastic scattering in a test beam} 

\author[a]{Giovanni Abbiendi,}										
\author[b,c]{Giovanni Ballerini,}										
\author[d]{Dipanwita Banerjee,}										
\author[d]{Johannes Bernhard,}										
\author[c,e,f]{Matteo Bonanomi,}										
\author[c,e]{Claudia Brizzolari,} 
\author[g]{Luca	G. Foggetta,}										
\author[k]{Mateusz Goncerz,}
\author[l]{Fedor V. Ignatov,}
\author[j]{Marco Incagli,}
\author[k]{Marcin Kucharczyk,}			
\author[a]{Umberto Marconi,}										
\author[b,c]{Valerio Mascagna,}										
\author[e]{Clara Matteuzzi,}
\author[j]{Riccardo Pilato,}
\author[m]{Dinko Pocanic,}
\author[b,c]{Michela Prest,}					
\author[a]{Antonio Principe,}
\author[b,c]{Federico Ronchetti,}										
\author[h,i]{Mattia Soldani,} 
\author[j]{Roberto Tenchini,}										
\author[c]{Erik Vallazza,} 
\author[j]{Graziano Venanzoni,}										
\author[k]{Mariusz Witek,}
\author[k]{Milosz Zdybal}	

\affiliation[a]{INFN Sezione di Bologna, \\Via Irnerio , Bologna, Italy}
\affiliation[b]{Università degli Studi dell'Insubria, \\Via Valleggio 11, Como, Italy}
\affiliation[c]{INFN Sezione di Milano Bicocca, \\Piazza della Scienza 3, Milano, Italy}
\affiliation[d]{CERN \\Esplanade des Particules, Geneva, Switzerland}
\affiliation[e]{Università degli Studi di Milano Bicocca, \\Piazza della Scienza 3, Milano, Italy}
\affiliation[f]{Laboratoire Leprince-Ringuet, CNRS/IN2P3, \\Ecole Polytechnique, Palaiseau, France}
\affiliation[g]{INFN Laboratori Nazionali di Frascati, \\Via Fermi 54, Frascati, Italy}
\affiliation[h]{Università degli Studi di Ferrara, \\Via Saragat 1, Ferrara, Italy}
\affiliation[i]{INFN Sezione di Ferrara, \\Via Saragat 1, Ferrara, Italy}
\affiliation[j]{INFN Sezione di Pisa, \\Largo Bruno Pontecorvo 3, Pisa, Italy}
\affiliation[k]{Institute of Nuclear Physics PAN, \\Ul. Radzikowskiego 152, \\Krakow, Poland}
\affiliation[l]{Budker Institute of Nuclear Physics, \\SB RAS, 11 Acad. Lavrentieva Pr., \\Novosibirsk, Russia}
\affiliation[m]{University of Virginia, \\Charlottesville, VA, USA}

\abstract{
In 2018, a test run with muons in the North Area at CERN was performed, running parasitically downstream of the COMPASS spectrometer.
The aim of the test was to investigate the elastic interactions of muons on atomic electrons, in an experimental configuration similar to the one proposed by the project MUonE, which plans to perform a very precise measurement of
the differential cross-section of the elastic interactions.

\noindent COMPASS was taking data with a 190 GeV $\pi$ beam, stopped in a tungsten beam dump: the muons from these $\pi$ decays passed through a setup including a graphite target followed by 10 planes of Si tracker  and a BGO crystal electromagnetic calorimeter placed at the end of the tracker.
The elastic scattering events were selected and analysed, and compared to expectations from MonteCarlo simulation. The agreement found was 
satisfactory and demonstrated that measuring the angles of the outgoing particles, a clean sample of elastic interaction could be identified.
}

\begin{document}

%
%
\maketitle 
\flushbottom 
\section{Introduction}
Recently a new experiment, MUonE, has been proposed with the aim of measuring the running of the effective electromagnetic coupling at low momentum transfer in the space-like region ($\alpha$(q$^2$), q$^2$ < 0) to provide an independent determination of the leading hadronic contribution to the (g-2)$_\mu$ of the muon \cite{original_paper}.
Such a measurement relies on the precise determination of the measured angles of the outgoing particles emerging from the elastic scattering 
$\mu+e \rightarrow\mu+e$ of  high-energy muons 
(160 GeV) impinging on atomic electrons of a light material (beryllium or carbon) target \cite{Muone2019}.

In 2018, a test run was performed at CERN, with a setup located behind the COMPASS spectrometer in the North Area \cite{compass07}, in order to set guidelines for the proposed configuration for MUonE.
The detector consisted of an 8 mm graphite target followed by a Si tracker and an electromagnetic calorimeter.
Despite the fact that the Si tracker used in this test had a worse (at least a factor 4) spatial resolution than the MUonE final apparatus \cite{Muone2019}, much interesting information was obtained from the data analysed in this paper.
 
\section{Experimental setup}


The 2018 test run was performed in EHN2, downstream of the COMPASS spectrometer, and exploited the ${\sim 187~\mbox{GeV}}$ positive muons that result from the decay of pions in the beam used by COMPASS. The remaining hadrons are stopped in a ${1.2~\mbox{m}}$ thick tungsten block located downstream of the COMPASS target or, when in muon beam mode, in nine ${1.1~\mbox{m}}$ thick beryllium modules that can be moved into the beamline via remote control ${\sim 400~\mbox{m}}$ upstream of COMPASS \cite{cern_beamlines_site,compass07}. 
The latter configuration was occasionally used for the COMPASS calibrations \cite{Muone2019}. 
At the location of the test setup, the muon beam has a width of several tens of centimetres. 

\paragraph{The tracker}
The core of the apparatus was the silicon microstrip-based tracking system developed by the INSULAb group \cite{Soldani_Talk2019}. 
Each of the 16 tracking planes consisted of a $9.293~\times~9.293~\times~0.041~\mbox{cm}^3$ single-side sensor with 384 channels, manufactured by Hamamatsu on high resistivity substrates for the AGILE experiment \cite{agile03}. 
The strips have a ${121~\mu\mbox{m}}$ pitch and, given the fact that the floating strip scheme is used, a ${242~\mu\mbox{m}}$ readout pitch. Each sensor is read out by three 128-channel, low-noise, analog-digital ASICs -- TA1 or TAA1 by IDEAS \cite{IDEAS_TA1, IDEAS_TAA1}. 
The readout of the differential analog output is multiplexed with a ${2.5~\mbox{MHz}}$ clock frequency. The analog readout allows an overall intrinsic spatial resolution of ${\sim 35~\mu\mbox{m}}$. 

\begin{figure}[htbp]
\centering 
\includegraphics[width=.65\textwidth]{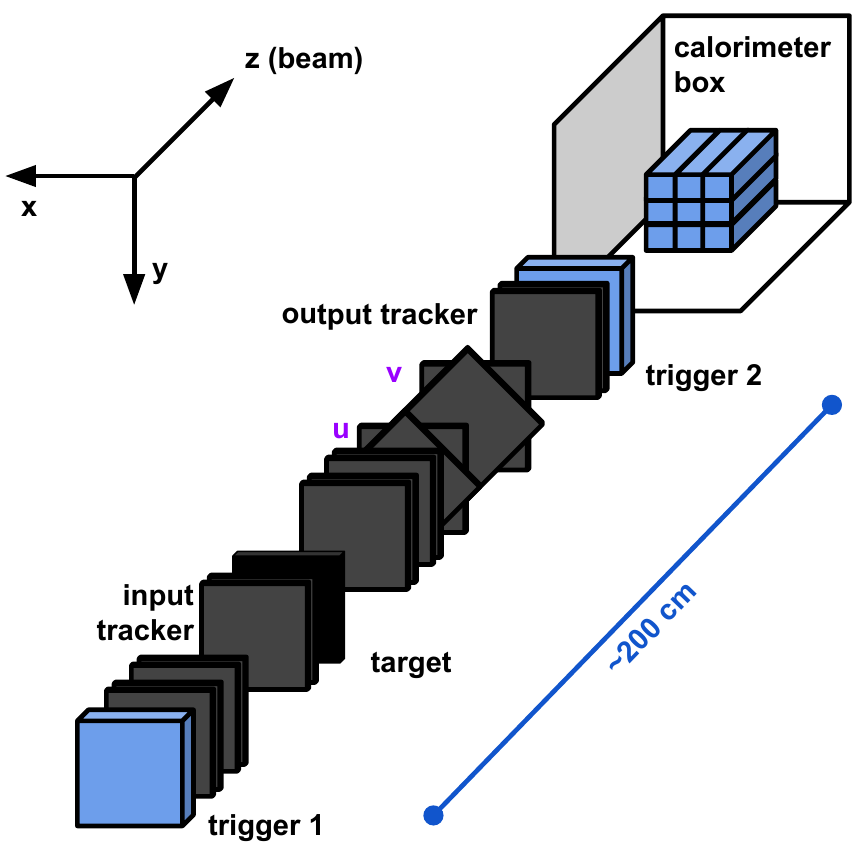}
\caption{Setup scheme: the gray boxes represent the silicon layers; the black box represents the graphite target; the blue boxes represent the scintillating layers.}
\label{fig:setup_scheme}
\end{figure}

Several geometrical configurations were tested, which differ from one another by the number of $8~\mbox{mm}$ thick graphite targets, the output tracker lever arm and the number of \textit{u} and \textit{v} ($\pi/4$) stereo layers. 
In the setup used for the present analysis 
6 (10) silicon planes were placed in front (behind) of a single target (see fig.~\ref{fig:setup_scheme}). 
The output stage was ${\sim 1.3~\mbox{m}}$ long, which resulted in a ${\sim 35~\mbox{mrad}}$ angular acceptance upper limit. 
Further details on the apparatus can be found in \cite{Soldani_Master2019}.
The target was a $10~\times~10$ cm$^2$ graphite layer, 8 mm thick.
It was installed on a custom plastic holder, coupled to the Newport rail via a Bosch mechanical support. 
\noindent The coincidence between the signals of two plastic scintillator trigger counters (with size $10~\times~10$~cm$^2$, shown in
fig.~\ref{fig:setup_scheme}), 
together with the beginning-of-spill and end-of-spill signals delivered by the SPS \footnote{
The typical SPS cycle for fixed-target (FT) operation lasts at least 14.8 s, including 4.8 s spill duration, i.e. the time during which the beam is slowly extracted. The number of FT cycles is about 2-3 per minute depending on LHC fillings and constraints by other users.}, allowed a clean trigger of muons passing inside the acceptance of the tracker.

\begin{figure}[htbp]
\centering 
\includegraphics[width=.6\textwidth]{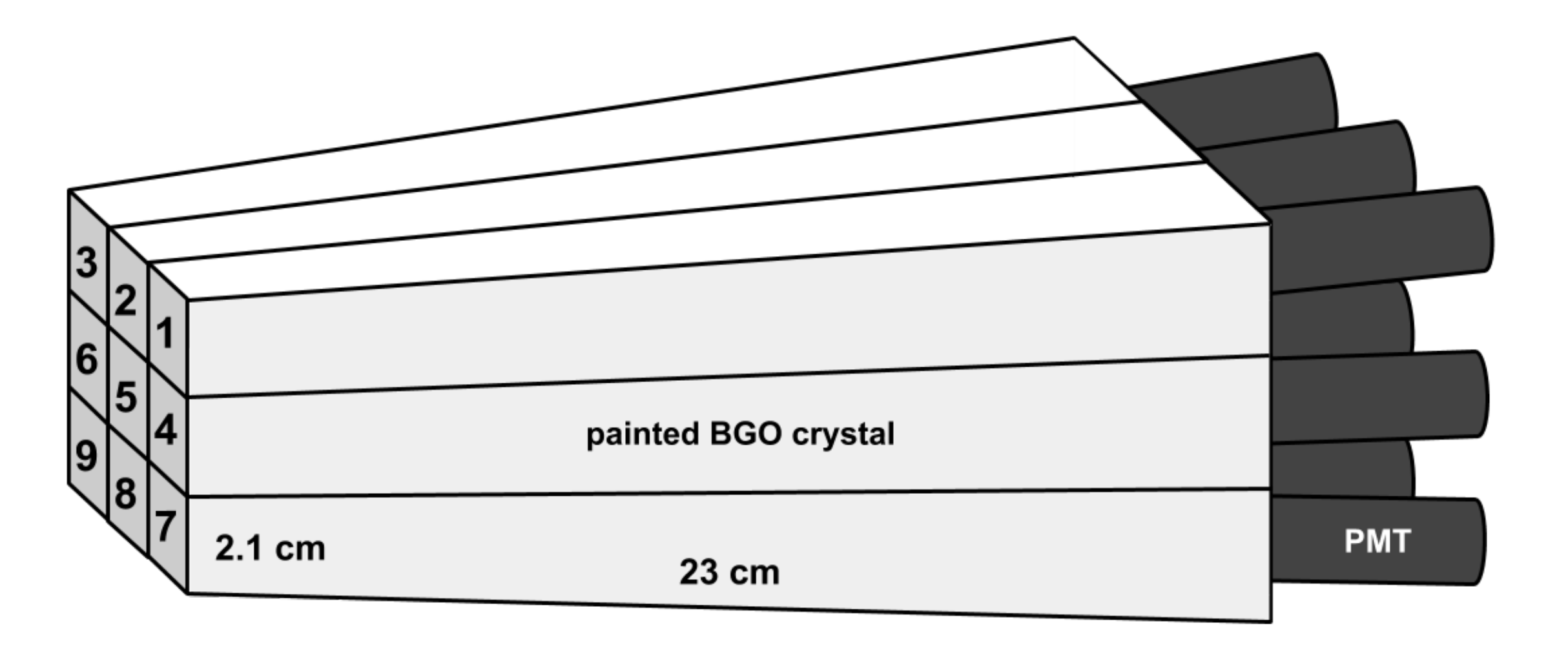}
\caption{ Scheme of the BGO calorimeter.}
\label{calo_GENNI}
\end{figure}

\paragraph{The electromagnetic calorimeter}
In the data set analysed in this paper, the calorimeter was a compact array of $3~\times~3$ BGO tapered crystals, each with a front (exit)  surface of ${\sim 2.1~\times~2.1 (2.8~\times~2.8)~\mbox{cm}^2}$ and
23 cm length. 
They were read out by Photonis XP1912 PMTs \cite{GENNI_PMTs} biased at $850~\mathrm{V}$.
The crystals were obtained by machining bigger spare blocks of the L3 endcap calorimeter \cite{L3} and were arranged as shown in fig. \ref{calo_GENNI}; such a configuration minimizes the dead regions in the detector active volume to the $\lesssim 1~\mathrm{mm}$ irreducible gaps introduced by the painting that shields and protects each block.

\begin{figure}[htbp]
\centering 
\includegraphics[width=.8\textwidth]{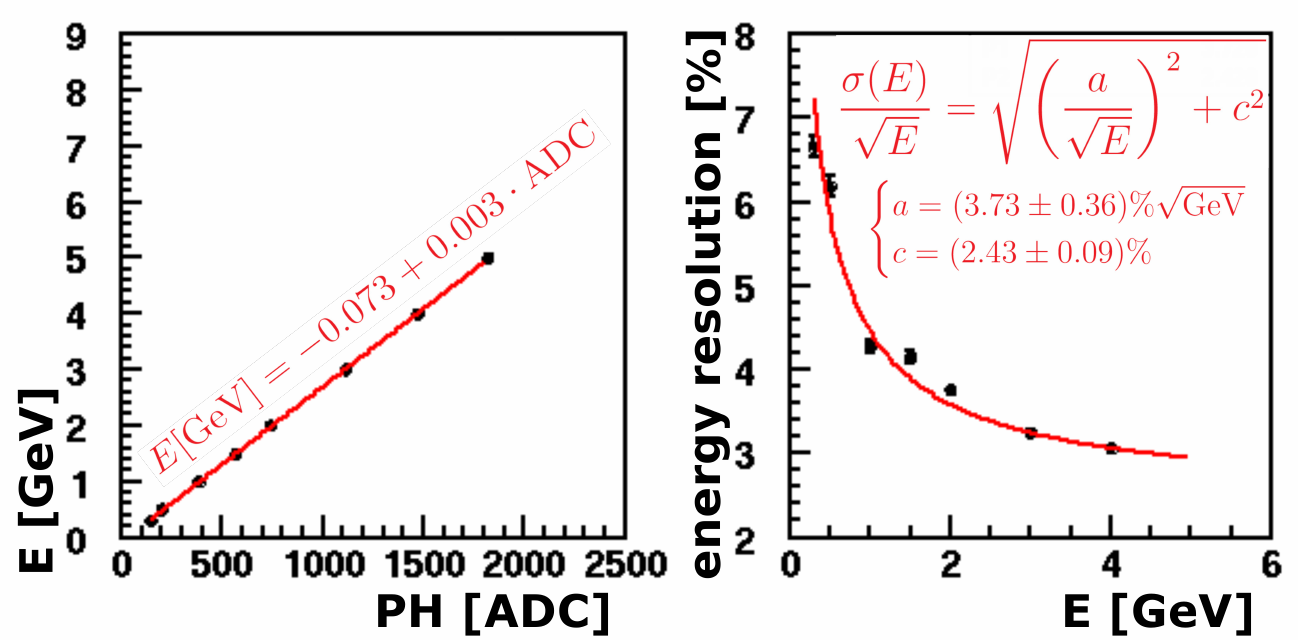}
\caption{ Calorimeter response linearity \textit{(left)} and energy resolution \textit{(right)} as a function of the detector signal pulse height (PH) \cite{Soldani_Master2019}. ADC stands for Analog-to-Digital Counts. The values of the fit parameters $a$ and $c$ are given in the text.}
\label{linearity_calo}
\end{figure}

The transverse size of the electromagnetic calorimeter covered an angular acceptance of about 15~mrad on each side from the center of a Si layer.
The detector performance in terms of linearity and energy resolution is shown in fig.~\ref{linearity_calo}. The measured energy resolution is 
$\sigma(E)/E = \left[ \right( a/\sqrt{E} \left)^2 + c^2 \right]^{1/2}$ with $a=(3.73\pm 0.36)\% \sqrt{\mathrm{GeV}}$ and $c=(2.43\pm 0.09)\%$. 
Further details on this calorimeter and on its characterization can be found in \cite{Soldani_Master2019}.

Given the finite range of the acquisition chain, when high-energy electrons impinge on the calorimeter, saturation may occur: in this case, the response of a single detector channel was capped at a maximum value, which corresponds to ${\sim 11~\mathrm{GeV}}$. 
This results in an overall upper limit in the measurement of the output electron energy of ${\sim 20~\mathrm{GeV}}$. 

\paragraph{The beam}
The data were taken parasitically while COMPASS was running with pions of 190 GeV energy.
The muons originated mainly from the decays of the pions stopped in the beam dump at the end of the COMPASS spectrometer.
The hadron content at the location of the test setup was completely 
negligible.

\begin{figure}[htbp]
\centering 
\includegraphics[width=.65\textwidth]{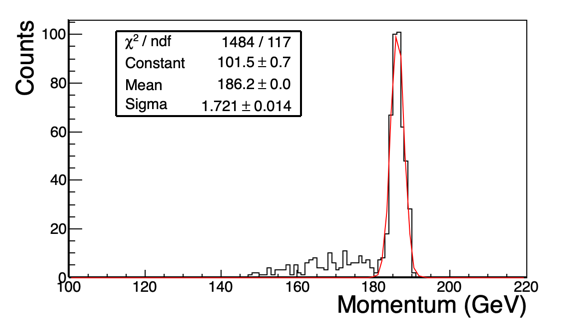}
\caption{Calculated energy profile of the muons beam reaching the test apparatus and originating from $\pi$ decays in the COMPASS dump. 
}
\label{beam_energy}
\end{figure}

\noindent The resulting energy profile of the muons entering the test apparatus is shown in fig.~\ref{beam_energy}, showing a peak at around 187 GeV with a tail. 
The divergence along the horizontal (vertical) direction is ${\sim 0.6~\mbox{mrad}}$ (${0.5~\mbox{mrad}}$),
with an intensity of ${\sim 0.6 \times 10^6}$ particles per spill.

\paragraph{Simulation} 
The MonteCarlo sample used for the analysis consists of 150'000 
$\mu$-$e$ elastic scattering events generated within the FairRoot \cite{FairRoot} framework and simulated using GEANT4 \cite{GEANT4}. 
The geometry and material properties of the detector used in 2018 test beam described above have been implemented in GEANT4, with the  simplification of defining a single-block calorimeter instead of the 9 crystals in the real setup. 
The distributions of the incoming beam $x$ and $y$ position have been chosen to match that of the reconstructed incoming tracks in data events with non-zero calorimeter deposit.

\noindent The incoming muon beam has been taken to be a monoenergetic beam of 187 GeV, the tail has not been considered in the simulation (c.f. fig.~\ref{beam_energy} ).

\noindent The $\mu$-$e$ events have been generated based on Leading Order (LO) calculations\footnote{The use of LO instead of NLO calculations, induces an uncertainty of $\sim$10\% on the measurement of the cross sections and the detection and reconstruction efficiencies. The effect on the shape of the observables, as done in this analysis, is smaller and depends on the variables and on the cuts applied \cite{Pavia:NLO}.}, and the track propagation and simulation have been done with GEANT4. 

\noindent Hits registered in the Si detectors were subsequently translated to their frame of reference and smeared by a Gaussian distribution with sigma corresponding to uncertainties determined from the measured data. 
The energy deposited by an event in the calorimeter corresponds to the sum of the deposits from all tracks passing through it.
\section{Event reconstruction and selection}
The operation of the test run lasted ${\sim 6}$ months. 
The analysis presented here concerns the data collected in the one-target configuration, 16 tracking planes
and the calorimeter shown in fig.~\ref{fig:setup_scheme} and described in the previous section.
The data were taken in the last period of the test beam operation.

\noindent After a first offline filter of the triggered events,
 a preliminary alignment was performed, and hits were required in the 6 tracking planes upstream of the target: this filter retained $\sim$2M triggers. 
More stringent requirements were then applied on the presence of an
incoming track and enough hits in the 10 planes after the target to allow the reconstruction of at least two tracks.
All these criteria reduced the sample to $\approx$ 94k events.





\paragraph{Alignment}

All the tracking layers, including stereo ones, were aligned based on the collection of good quality reconstructed tracks with at least ten hits. 
The $(x,y)$ position of the first layer was taken as a reference. 
The shift in $(x,y)$ plane and the rotation angle of other layers around the $z$ axis were determined with respect to the reference layer. 
An iterative procedure was applied. 
The $z$ positions of layers were taken from the measured values of a geometrical survey. 
In one iteration the layers were aligned one-by-one. 
A track was refitted excluding a given layer and the sum of the residuals of all tracks from the collection was minimized with respect to $x$ and $y$ shift and rotation angle of the excluded layer. 
The iteration was finished when the change in the parameters of all layers was below a given threshold. 
The distributions of the residuals obtained for final parameters were then fitted using a single Gaussian distribution to determine the resolutions of individual layers. 
The resolutions  varied from 15 to 37 $\mu$m, with the 
spread mainly due to the intrinsic quality of the sensors and the readout chain.

\paragraph{Tracking algorithm}

The scattering of high-energy muons on atomic electrons of a low-Z target through the elastic process is characterized by a simple topology. 
Three tracks are expected to be reconstructed in the detector, i.e. the incident muon before the target and outgoing electron and muon after the target. 

\noindent The track reconstruction is performed separately in detector parts before and after the target. 
First, the two-dimensional (2-d) tracks are searched for independently in $x$-$z$ and $y$-$z$ projections. In the next step, accepted 2-d track candidates are combined into three-dimensional (3-d) tracks. 
Then the track fit is performed including hits in the stereo layers. 
The elastic $\mu$-$e$ scattering event is obtained from reconstructed 3-d tracks: the track reconstructed before the target and the two tracks reconstructed after the target are checked for compatibility to belong to the common interaction vertex. 
The interaction vertex is constrained to the center plane of the target. 

\paragraph{Track reconstruction}
The 2-d track finding is performed in each projection by constructing pairs from all the combinations of hits in $x$ and $y$ layers separately. 
For each pair of hits, a 2-d line in $x$-$z$ or $y$-$z$ projections is determined. 
To maximize the efficiency, all the hits compatible with the straight line within a relatively wide window corresponding to 10 times the sensor resolution are collected. 
A fit is then applied to the selected combinations, after removing outliers. 
The set of 2-d track candidates is sorted according to the number of collected hits and the $\chi^{2}$ of the fit. 
In the last phase a clone killing procedure is applied as follows:
only the best tracks with unique combinations of hits are accepted. 
At least 3 hits in each projection are required.
All pairs of track candidates, in $x$-$z$ and $y$-$z$ projections, are combined into 3-d track candidates. 
The compatible hits from stereo layers are included and the track fit is performed. 
An iterative fitting procedure is applied using the least square method.  
After each iteration, hits more than 5$\sigma$ away from the fitted line are removed, and the  fit is repeated until no outlier is found. 
As no unique combination of hits forming 3-d lines is imposed up to this point, the collections of tracks may contain clones, where clones are defined as those track candidates containing common hits. 
The clone removal procedure is applied as follows: the tracks are
sorted according to the number of hits and $\chi^2$ per number of degrees of freedom NDF of the least square fit ($\chi^2/NDF$). 
The tracks with the largest number of hits are accepted first. For the same number of hits the candidate of best quality is taken using the $\chi^2/NDF$ criterion. 
After accepting a track, the hits used by that track are not considered any further. 
Then the next track from the sorted list is searched for and accepted if it contains the required number of hits (3 hits in $x$ projection and 3 hits in the $y$ projection) not used by any track already accepted.
The final collection of tracks with unique set of hits is passed to the last stage of the event reconstruction.

\paragraph{Reconstruction of $\mu$-$e$ scattering events}
The set of reconstructed tracks is used to search for events with the elastic $\mu$-$e$ scattering topology. In the first step all combinations
of track pairs reconstructed after the target are checked to be compatible 
with intersecting inside the target. 
Then a third track, incoming to the target and passing close to the intersection point, is searched for. 
For the three tracks initially compatible with muon-electron scattering, a dedicated vertex fit is performed to obtain the best possible accuracy for the scattering angles of the outgoing muon and electron. 
To take into account multiple scattering, the momentum of tracks has to be estimated. 
For the small ($<2.5$ mrad) scattering angles of both outgoing muon and electron, the momenta are expected to be large enough to neglect multiple scattering in the material. 
For angles above 2.5 mrad, the track with the largest angle is assumed to be the electron. 
If one assumes that the selected three-track event corresponds to a genuine $\mu-e$ elastic scattering, the observed scattering angle of the electron can be used to estimate its momentum. 
The expected value of the momentum is assigned to the electron candidate
using the knowledge of the beam momentum and of the two-body kinematics.
The other track from the pair after target is assumed to be the muon. 
For such outgoing muons the expected momentum is high enough to neglect multiple scattering, defined as described above. 
A dedicated kinematic fit of the vertex is performed, based on a constrained least square method. 
The common ($x_{vtx}$,$y_{vtx}$) position, at the middle $z$ coordinate of the target, is enforced.
 The uncertainties of the hits assigned to the electron track are estimated using the predicted multiple scattering. 
 The uncertainties due to detector resolutions and multiple scattering, from all material from target up to the $z$ position of a given hit, are added in quadrature neglecting the correlations.
 The least square fit uses the 3-d line slopes of the three tracks and ($x_{vtx}$,$y_{vtx}$) as free parameters. 
The total $\chi^2$
used for minimization is the sum of the $\chi^2$ contributions from all hits of the three tracks.
The total vertex $\chi^2$/NDF, referred to as $\chi^2_{vtx}$, will be used through the paper.  
Its distribution is shown in fig.~\ref{acoplanarity_chi2-10}(a).   

\begin{figure}[htbp]
\includegraphics[width=.45\textwidth]{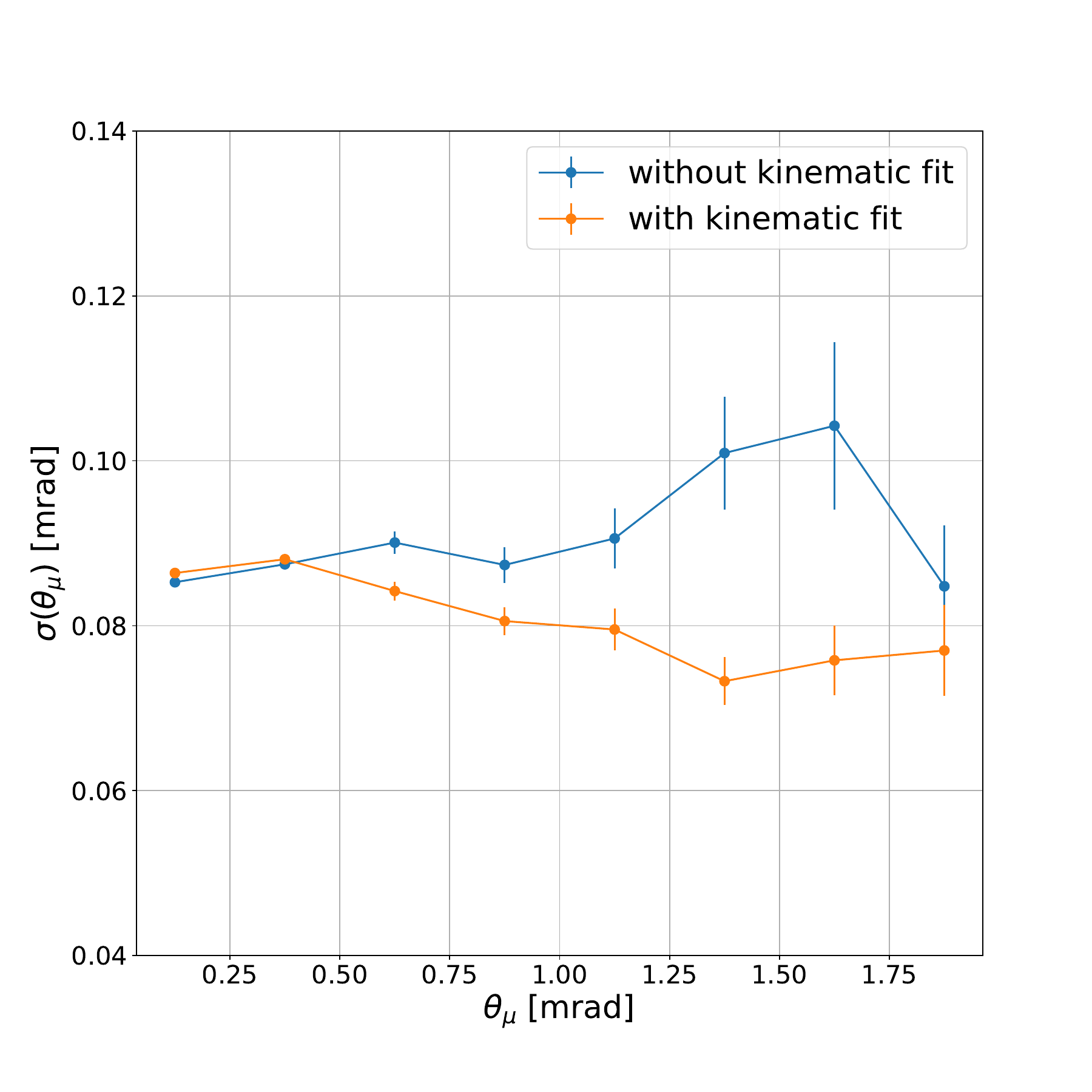}
\includegraphics[width=.45\textwidth]{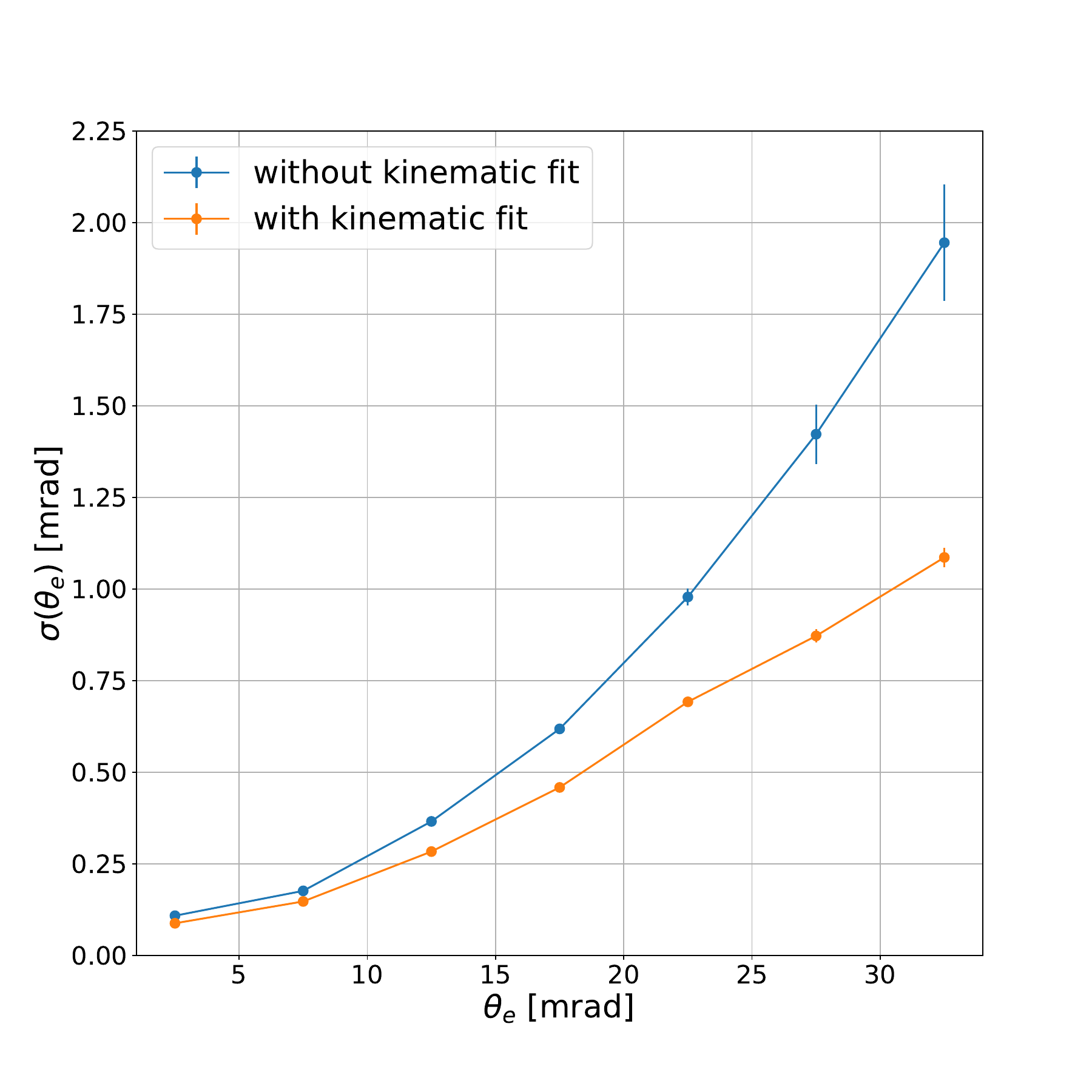}
\caption{Angular resolution as a function of the scattering angle for ($left$) muons and for ($right$) electrons, determined with simulated events. 
The curves correspond to 
before (in blue) and after (in yellow) the dedicated kinematic fit of the $\mu$-$e$ scattering vertex described in the text. 
}
\label{angular_resolution}
\end{figure}

The angular resolution for the two outgoing tracks is then determined from the MonteCarlo simulation as the $\sigma$ of the Gaussian function fitting the difference between the true angle and the reconstructed angle, plotted in 
fig.~\ref{angular_resolution} as a function of the true emission angle.
For muons it turns out to be quite flat around 0.080 mrad, while for electrons it varies significantly as a function of the angle (i.e. the energy) from 0.100 to 0.900 ~mrad, mainly due to multiple scattering. 

The sample of $\sim$ 94k events is reduced to $\sim$ 56k events after the final alignment and  by requiring 
only one incoming track, a total deposit in the calorimeter > 0, and enough hits in the tracker to reconstruct at least 2 tracks in the final state.
Once the full reconstruction was performed, requiring at least three hits per plane and per track, and fitting a common vertex, 8556 events remain.


\paragraph{Selection of $\mu$ - $e$ scattering events}
Two loose initial cuts are applied at the first stage of the analysis and will be included in all further cuts described in this paper: 1) the $\chi^2_{vtx} <$  10 cut, determined from fig.~\ref{acoplanarity_chi2-10}(a),
and 2) a cut on the electron emission angle rejecting events with $\theta_e > 30$ mrad. This last cut is applied  in order to 
suppress low energy electrons, considering that the simulated events are generated with E$_e$> 1 GeV, which implies an angular cut at $\theta_e \approx 35$~mrad. 
The comparison of the data with the simulated elastic events traced through the experimental setup with GEANT4 is then valid in first approximation.

\noindent The discrepancy between data and simulation is due to the presence of background in the data. The simulation contains only a pure sample of elastic scattering events. 

The effect of these initial loose cuts is shown in the first three rows of table~\ref{tab:event_data}.

\section{Analysis and results}

\begin{figure}[htbp]
\includegraphics[width=.45\textwidth]{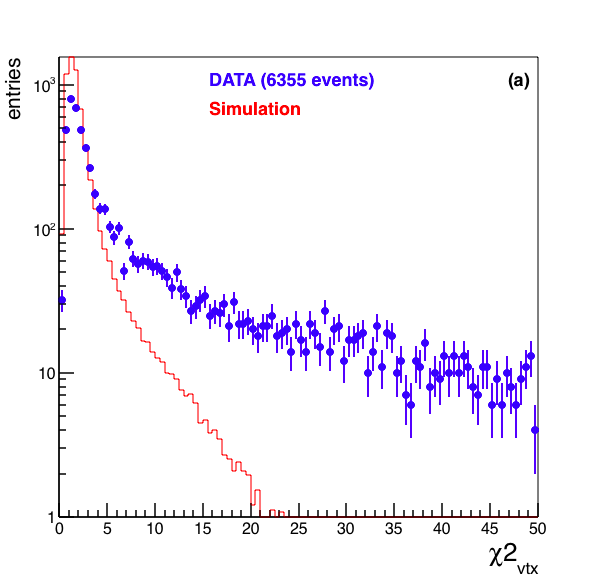}
\includegraphics[width=.50\textwidth]{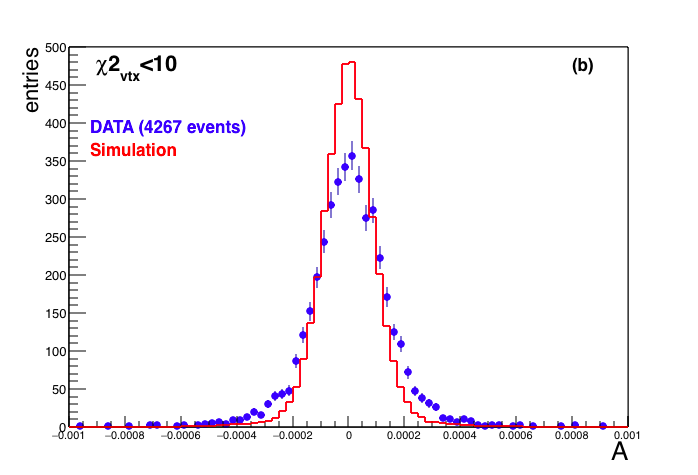}
\caption{(a):$\chi^2_{vtx}$ distribution. (b): acoplanarity distribution. Data are in blue with error bars, the simulated and reconstructed elastic events are the red histogram and are normalized to the observed number of events in data.}
\label{acoplanarity_chi2-10}
\end{figure}

The specific kinematics of elastic scattering events requires that the events are planar, and that the two angles of the outgoing $\mu$ and $e$ are strongly correlated.
The acoplanarity ($A$), defined as the angle between the incoming muon and the plane of the two outgoing particles,  
is shown in fig.~\ref{acoplanarity_chi2-10}(b) for measured data and simulated elastic events, after the cut at $\chi^2_{vtx} < 10$.

\begin{figure}[htbp]
\includegraphics[width=.45\textwidth]{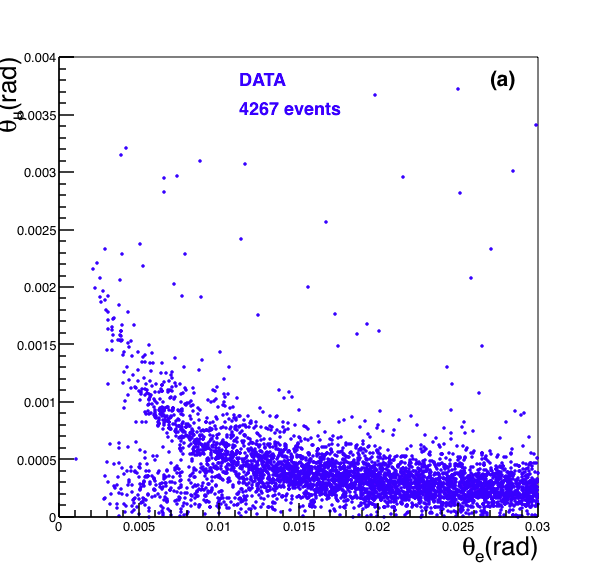}
\includegraphics[width=.45\textwidth]{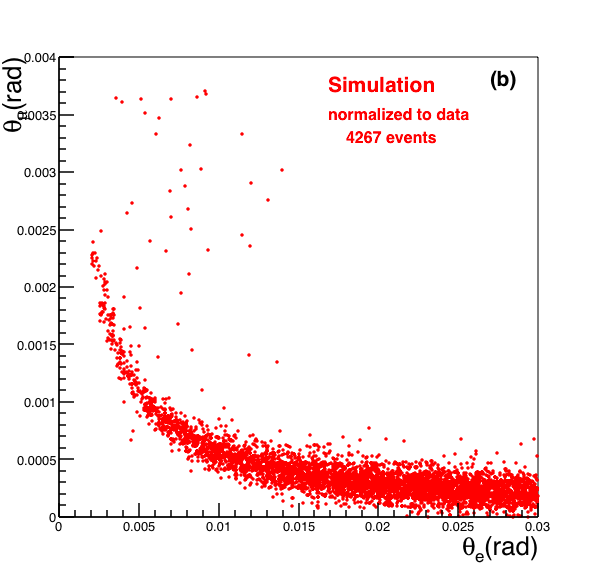}
\caption{(a) kinematical correlation for data and (b) for simulated elastic events.}
\label{th_e-th_mu_chi2-10}
\end{figure}

\noindent The kinematical correlation for ($\theta_\mu,\theta_e$) is shown in fig.\ref{th_e-th_mu_chi2-10} at this stage of the selection.
In the measured data (fig.\ref{th_e-th_mu_chi2-10}(a)) 
there is clearly the contribution from background, visible outside the correlation curve.

\begin{figure}[htbp]
\centering 
\includegraphics[width=.60\textwidth]{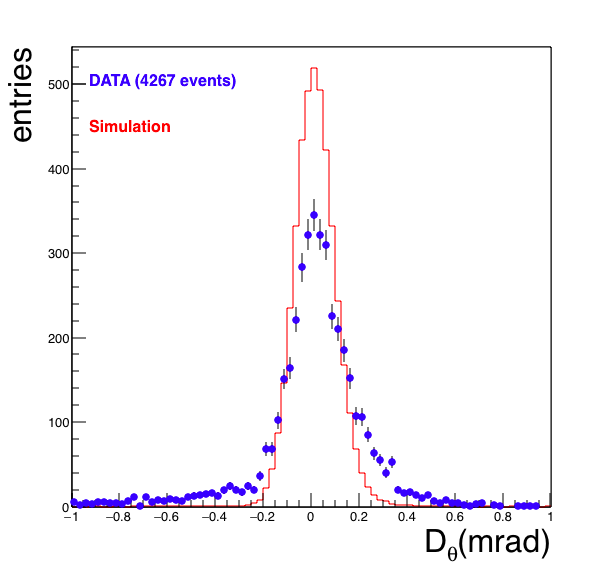}
\caption{$D_\theta$, defined in the text, for data and simulated elastic events normalized to the number of events observed. 
}
\label{Dtheta_chi2-10}
\end{figure}

\begin{figure}[htbp]
\centering 
\includegraphics[width=.60\textwidth]{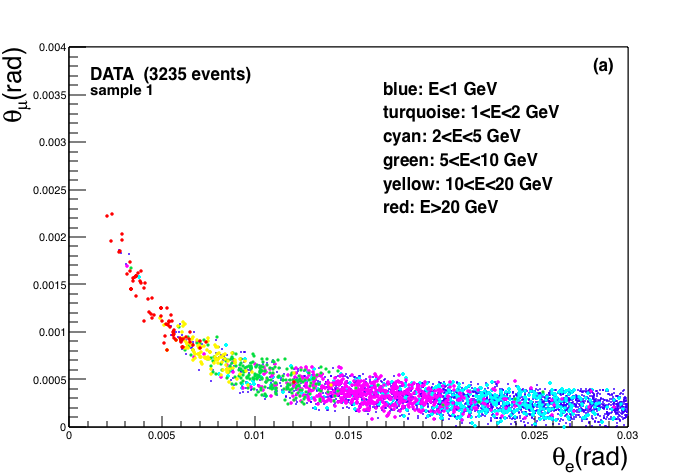}
\includegraphics[width=.60\textwidth]{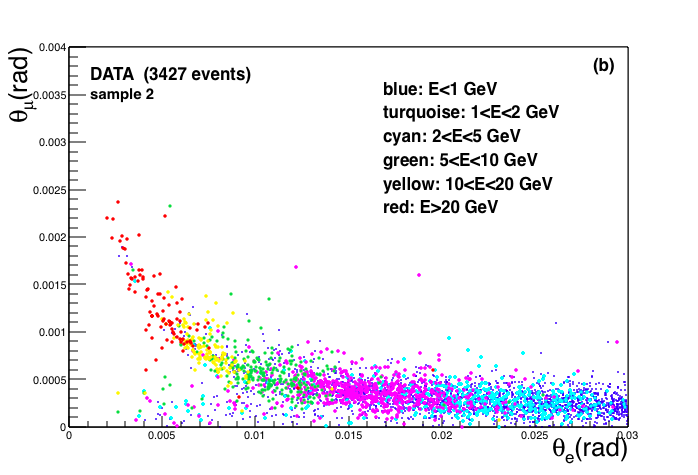}
\caption{Kinematical correlation of the outgoing muon and electron, with the color code representing the energy deposited in the calorimeter. (a) for measured data after the cuts on $\chi^2_{vtx} < 5$ and
$|A|<3.5\cdot 10^{-4}$ (sample 2 in table~\ref{tab:event_data}), and (b) the measured data after the cut on $|D_\theta| < 0.2$ mrad (sample 1).
}
\label{kin_corr}
\end{figure}

 A variable $D_\theta$, which estimates  the elasticity of 
a reconstructed event, is defined as 
the minimum angular distance of the measured event
 to the expected theoretical kinematic curve, and
 is calculated for a given incoming muon beam energy. 
This variable was introduced and used in the NA7 experiment \cite{NA7} to reject and estimate backgrounds.
The distribution of $D_\theta$ is shown in 
fig.~\ref{Dtheta_chi2-10} for data and for simulated elastic events.
The discrepancy between data and simulation is due to the presence of
background in data, while the simulation contains only elastic scattering events.
Based on the simulation, a cut was set at $-0.2 < D_\theta<0.2$ mrad.
After this cut is applied, 3235 events remain, and their kinematic correlation is shown 
in fig.~\ref{kin_corr}(a), where the information on the deposited energy in the calorimeter for these events is also shown,   
which approximately corresponds to the energy of the electrons \footnote{The calorimeter structure doesn't allow the separation of muon and electron deposited energies.}.

For comparison, an alternative set of selection criteria was applied, requiring $\chi^2_{vtx} < 5$ and |$A$| < $3.5 \times 10^{-4}$, and no cut on $D_\theta$. 
This set of cuts yields 3427 events.
For this sample the data plot of ($\theta_e,\theta_{\mu}$) is shown in  fig.\ref{kin_corr}(b), containing also the information on the deposited energy in the calorimeter.
 

\noindent The summary of event yield at each selection step is given in table~\ref{tab:event_data}, where the definitions of the two final data samples, sample 1 and sample 2, are given.

\begin{figure}[htbp]
\centering 
\includegraphics[width=.60\textwidth]{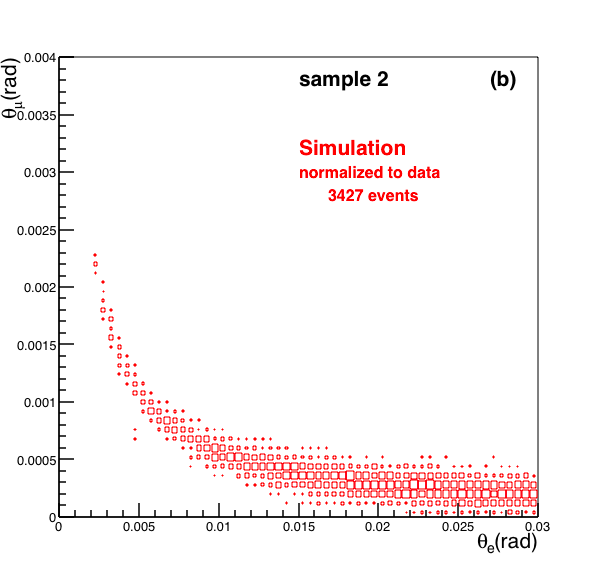}
\caption{
The kinematical correlation 
 for sample 2 (defined in table~\ref{tab:event_data}):
as expected from simulated elastic scattering events. }
\label{kin_corr_standard-cuts}
\end{figure}

\begin{table}[ht!]
  \begin{center}
    \caption{Event yields after each step in the selection.
    The bottom two rows are alternatives, defining two different samples, labelled sample 1 and 2,  used in the analysis.
    }
    
    \label{tab:event_data}
    \begin{tabular}{|c|c|}
    \hline
    \bf{Selection criteria} & \bf{number of events}   \\
    \hline
    Initial sample   	  & 8556 \\
    \hline
    $\theta_e<30$ mrad  & 6355	\\
    \hline
    $\chi^2_{vtx} <10$     & 4267	\\
    \hline
    \bf{Above criteria and:} &    \\
    \hline
    sample 1: |D$_\theta| < 0.2$~mrad              & 3235 \\
    \hline
    sample 2: $\chi^2_{vtx} < 5$ and |A|$<0.00035$  & 3427 \\
\hline
\end{tabular}
\end{center}
\end{table}



The selected events have been compared with elastic events
generated with LO cross section and simulated with GEANT4.

The correlation plot resulting from the selection of sample 2 is 
shown in fig.~\ref{kin_corr}(b).
Comparing with what expected for simulated elastic scattering events
(see fig.~\ref{kin_corr_standard-cuts}) it is clear that a residual background is visible in the data below the correlation curve.
This background is eliminated in 
sample 1 (c.f. fig.~\ref{kin_corr}(a)). 

The samples selected with different sets of cuts contain clearly 
different background sources (c.f. fig.~\ref{kin_corr}(a)
and fig.~\ref{kin_corr}(b)).
This suggests that, once identified with the proper simulation
the $\mu$ interactions responsible for the background, one can study 
with the data themself the level and shape of the residual contamination affecting the final elastic selected candidates.

\begin{figure}[htbp]
\includegraphics[width=.50\textwidth]{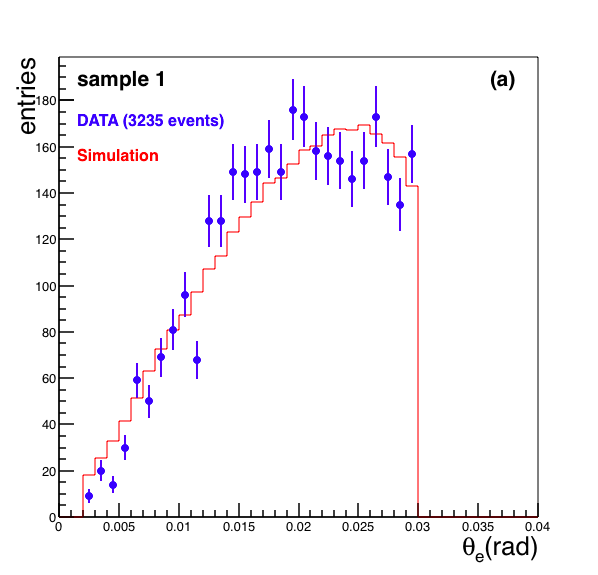}
\includegraphics[width=.50\textwidth]{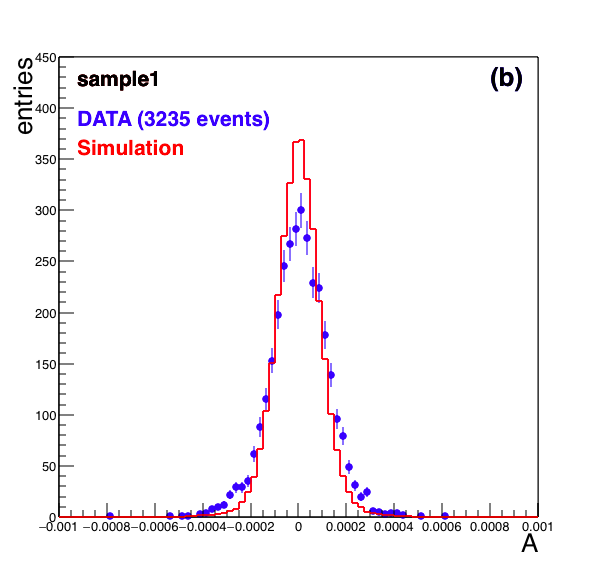}
\caption{Comparison data/MonteCarlo for the electron emission angle (a) and the acoplanarity (b). The simulated events have been normalized to the number of observed events. Sample 1 is defined in 
table~\ref{tab:event_data}.}
\label{DATA-MC_electronTheta}
\end{figure}

The reconstructed electron angle $\theta_e$ is shown in 
fig.~\ref{DATA-MC_electronTheta}(a) and the acoplanarity in 
fig.~\ref{DATA-MC_electronTheta}(b) for sample 1.

\begin{figure}[htbp]
\includegraphics[width=.45\textwidth]{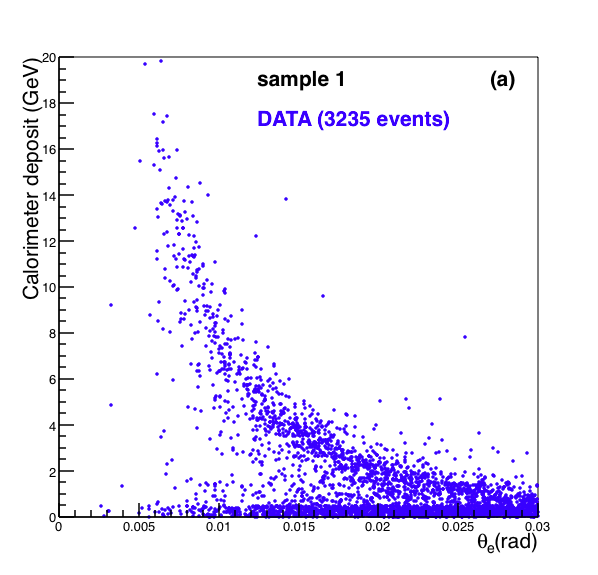}
\includegraphics[width=.45\textwidth]{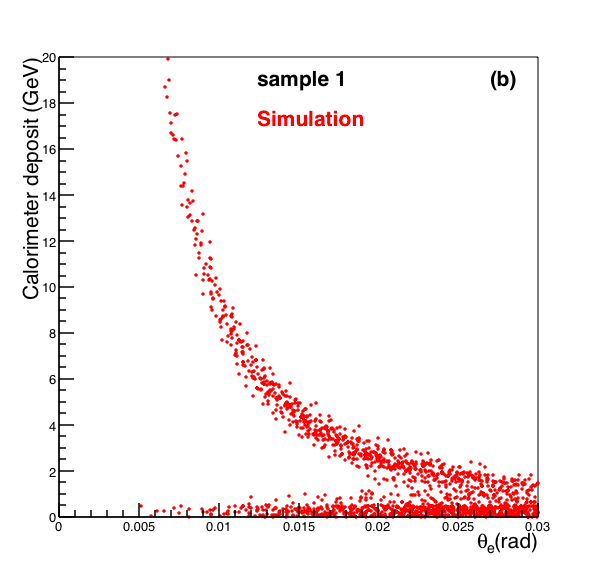}
\caption{Correlation between the reconstructed angle $\theta_e$ and the deposited energy in the calorimeter, assumed to be $E_e$ for measured data (a). For simulated events (b), in the MonteCarlo no detailed simulation is done for the calorimeter and the energy corresponds to the true energy deposited in the crystals.
}
\label{elec_angle-vs-energy}
\end{figure}

\noindent The correlation between the electron energy $E_e$ and $\theta_e$ is shown in fig.~\ref{elec_angle-vs-energy}.
The measured correlation is well described by the simulation.
The band of events in fig.~\ref{elec_angle-vs-energy}
with $E_e < 1$ GeV is explained by the combination of the small angular acceptance of the calorimeter (15 mrad from the center of the Si tracker), and the flat spacial profile of the incoming beam. 
When the electron is emitted at {\it{large}} angle it misses the calorimeter partially or completely; when it is emitted at {\it{small}} angle 
($ <15$ mrad) it partially or completely misses the calorimeter depending on wether the production vertex is on the border of the Si sensor. This correlation is clearly confirmed by the simulation.  

\noindent The comparison data-MC shows reasonable agreement, and 
fig.~\ref{DATA-MC_electronTheta}(b)
shows that some small background is surviving in the signal region.
One can use the side bands of the D$_\theta$ distribution (\textit{cf.} fig. \ref{Dtheta_chi2-10}) to roughly estimate it.
 The main contribution comes from below the kinematical curve, and it contains radiative events, pair production from the muon track,
 and migration of events from the region $\theta_e>30$ mrad.
 The events above the theoretical curve come mainly from badly measured events (i.e. events with a high $\chi^2_{vtx}$ value).

\noindent Using only the left side events, the extrapolation to the signal region $-0.2< D_\theta<0.2$ mrad is made using a Gaussian function describing the left tail of the distribution.
The area of this Gaussian in the range |D$_\theta$|$<0.2$ mrad is taken to define the background level.
The background under the signal peak turns out to contain of the order of 4\% of events, with an uncertainty around 50\%.

\noindent The use of the simulation of background channels, like $\mu e^+e^-$, will provide the right way to understand the level and the shape of this background.\\
To this end, the implementation of a more detailed and precise  simulation of the muon electromagnetic interactions in GEANT4 
is necessary.

Finally, we compare the ratio of number of events in two different angular regions after the different selection cuts.
The two angular regions are defined as 
$\theta_e < 5$ mrad and $15 < \theta_e < 20$ mrad.
These regions are chosen in view of the fact that, in a dedicated high precision experiment \cite{Muone2019}, to measure the hadronic contribution to the running of $\alpha$, the small angular region will be where these corrections would appear. Given the low event statistics collected in this test beam measurement, present results provide only preliminary and rough information. 
The comparison of measured event yields to LO MonteCarlo simulation is given in table~\ref{ratios}.
The large statistical uncertainties in the measured data make it difficult and premature to draw conclusions regarding the level of agreement.  


\begin{table}[ht!]
\centering

\begin{tabular}{|c|c|c|}
\hline
\multicolumn{1}{|c|}{$\frac{\theta < 5~\mathrm{mrad}}{\theta = 15-20~\mathrm{mrad}}$} & \multicolumn{1}{c|}{\begin{tabular}[c]
{@{}c@{}}$\chi^2 < 5$ \\ $|A| < 0.00035$\end{tabular}} & \multicolumn{1}{c|}{\begin{tabular}[c]{@{}c@{}}$|D_{\theta}|<0.2$ mrad \end{tabular}}  \\
\hline

\multicolumn{1}{|c|}{DATA} 
& \multicolumn{1}{c|}{\begin{tabular}[c]{@{}c@{}}$0.08 \pm 0.01$ \end{tabular}} & \multicolumn{1}{c|}{\begin{tabular}[c]{@{}c@{}}$0.06 \pm 0.01$ \end{tabular}} \\
\hline
\multicolumn{1}{|c|}{MC LO} & 
\multicolumn{1}{c|}{\begin{tabular}[c]{@{}c@{}}$0.100 \pm 0.002$ \end{tabular}} & 
\multicolumn{1}{c|}{\begin{tabular}[c]{@{}c@{}}$0.104 \pm 0.002$ \end{tabular}} \\

\hline
\end{tabular}

\caption{ Ratios of number of events measured in two different angular regions, satisfying two sets of selections cuts, as indicated. The comparison is made with LO simulated $\mu$-$e$ elastic scattering events.
Errors are statistical only.
}
\label{ratios}
\end{table}

\section{Conclusion}
In a test beam measurement performed parasitically behind the COMPASS spectrometer in the CERN North Area, elastic scattering $\mu$-$e$ interactions were studied. 
This preliminary investigation was aimed mainly to explore the ability to select a clean sample of elastic scattering events in view of designing
an experiment to measure the hadronic contribution to the running of $\alpha$.
The experimental test setup had a resolution worse than the one planned to be used in MUonE \cite{Muone2019}, but even in these conditions, we were able to select a clean sample of elastic events.

\noindent Several other running conditions were different in this test with respect to the planned MUonE conditions, such as the high intensity and the shape of the beam requested by MUonE, and therefore some experimental aspects could not be adressed.

This study however suggests the importance of an adequate  calorimeter, to understand the electrons emitted in the range of a few GeV, and the determination and behaviour of the background.

A crucial point for a future precise measurement of the differential cross section of the elastic $\mu$-$e$ process is the  
upgrade \cite{vladimir} of GEANT4, at present under test.
The upgrade concerns the muon pair-production interactions $\mu \rightarrow \mu ee$ for which an accurate angular distribution of the electrons of the pair has been implemented.
This upgrade is available in version 10.7 of the GEANT4 package, currently in the process of being validated.


%
\acknowledgments

We warmly thank the COMPASS collaboration for their kind willingness to
let us run parasitically behind their detector,
and in particular Vincent Andrieux for his active help.
We also thank the LNF-BTF and the PADME collaboration for giving us the BGO calorimeter, and the LNF-SPCM, where Tommaso Napolitano and Fabrizio Angeloni provided the mechanical support structure.

\bibliography{bibliography} 
%
%

\end{document}